# Optimize electron beam energy toward in-situ imaging of large thick bio-samples with nanometer resolution

## Xi Yang[1]*, Victor Smaluk[1], Timur Shaftan[1]

*[1]National Synchrotron Light Source II, Brookhaven National Laboratory, Upton, NY 11973, USA*
***Corresponding authors: xiyang@bnl.gov***

## ABSTRACT

To optimize electron energy toward in-situ imaging large bio-samples up to 10-μm thickness with nanoscale resolution, we implemented an analytical model based on elastic and inelastic characteristic angles [1]. This model can be used to predict the transverse beam size broadening as a function of electron energy while the probe beam traverses through the sample. As result, the optimal choice of the electron beam energy can be realized. While the sample thickness is less than 10 μm, there exists an optimal electron beam energy below 10 MeV regarding a specific sample thickness. However, for samples thicker than 10 μm, the optimal beam energy is 10 MeV, and the ultimate resolution could become worse with the increase of the sample thickness.

## INTRODUCTION

Driven by life-science applications, high energy mega-electron-volt Scanning Transmission Electron Microscope (MeV-STEM) could potentially break the fundamental limitation set by low-energy Electron Tomography (cryo-ET): the uncertainty and slow speed of Cryo-Focused Ion Beam slicing of large biological samples. This technique can produce just a few 300-nm-thick lamellae per day, and it often takes a few days to obtain an intact 3D biological cell image. Elastic and inelastic scattering of high energy ($\geq 10\ MeV$) electrons with the unique combination of small characteristic angle and high penetration could turn the amplitude-contrast MeV-STEM into an appropriate microscope for sample thickness up to 10 $\mu m$ with many applications in chemistry, biology, and life-science [1]. However, to minimize the geometrical broadening (GB) due to sample thickness and electron beam divergence, the probe beam must be focused on the specimen with a nanometre size and a milliradian semi-convergence angle. For an MeV-STEM instrument, a high-energy (3-10 MeV) and high-brightness (a few picometer geometrical emittance, <10⁻⁴ relative energy spread, and 1 nA beam current) electron source is essential. Our recent progress on the MeV-STEM design[1] shows a beyond state-of-the-art electron source can be realized via two different approaches: 1) DC gun, aperture, superconducting radio frequency (SRF) cavities, and condenser lens; 2) CW-SRF gun, aperture, SRF cavities, and condenser lens. Moreover, to mitigate the plural effects degrading the spatial resolution for large thick bio-samples, the electron beam energy must be boosted to 10 MeV or higher. As a result, despite where the electron beam being focused along the sample thickness dimension (e.g., top, middle, and bottom), the transverse size of the electron beam when it traverses through the sample can be kept $\leq 10\ nm$; thus, the size of the projected probe electron column in the STEM imaging mode can be minimized.

To achieve a resolution better than 10 nm for in-situ imaging of large bio-samples with the thickness up to 10 μm, we have implemented an analytical model based on the elastic and inelastic characteristic angles. This model can be used to predict the transverse beam size broadening as a function of electron energy. To keep the beam size below 10 nm along its path in the sample, the electron energy is required to be 10 MeV or more. Determining the ultimate resolution, especially for thick bio-samples, is a complex problem, which is beyond the scope of this manuscript. Also, further increasing the electron beam energy couldn't improve the overall resolution; instead, it just increases the cost and technical complicity of the instrument. Hence, the optimal selection of the electron beam energy for in-situ imagining of large thick bio-samples with a thickness up to 10 $\mu m$ is 10 MeV.

# RESULTS

## I. Analytical Model based on Electron Cross Sections

The angular distribution of scattering from a target atom can be described by differential scattering cross-section. In the Wentzel approximation, the differential cross-section for elastic scattering in the first-order Born approximation becomes [2]:

$$\frac{d\sigma_{el}}{d\Omega} = \left[\frac{2ZR^2\left(1+\frac{E}{E_0}\right)}{a_H(1+(\frac{\theta}{\theta_{el}})^2)}\right]^2 \;,\; \theta_{el} = \frac{\lambda}{2\pi R} \;,\; R = a_H Z^{-1/3} \tag{1}$$

where $\sigma_{el}$ is the total elastic scattering cross-section, $\Omega$ is the solid angle, $Z$ is the atomic number, $E$ is the electron energy, $E_0$ is the rest energy of the electron, $a_H$ is the Bohr radius (0.0529 nm), $\theta$ is the scattering angle, $\theta_{el}$ is the characteristic angle below which 50% of the electrons are elastically scattered into, $\lambda$ is the electron wavelength. Integrating equation (1) yields the total cross-section [3]:

$$\sigma_{el} \approx \frac{h^2 c^2 Z^{4/3}}{\pi E_0^2 \beta^2} \;,\; \beta^2 = 1 - \left[\frac{E_0}{E+E_0}\right]^2 \tag{2}$$

The angular dependence and the cross-section of inelastic scattering can be approximated with a Bethe-model [3,4]:

$$\frac{d\sigma_{inel}}{d\Omega} \approx \frac{Z\lambda^4\left(1+\frac{E}{E_0}\right)^2}{4\pi^2 a_H^2}\left[\frac{1-\left(1+\frac{\theta^2}{\theta_E^2}\right)^{-2}}{(\theta+\theta_E^2)^2}\right] \;,\; \theta_E = \frac{\Delta E}{E}\frac{E+E_0}{E+2E_0} \tag{3}$$

where $\sigma_{inel}$ is the inelastic scattering cross-section, $\theta_E$ is the angle determining the decay of the inelastic scattering, $\Delta E$ is the mean energy loss from a single inelastic scattering event (e.g., 39.3 eV for amorphous ice [5]). An inelastic scattering is concentrated within much smaller angles than elastic scattering. Identically, we define $\theta_{inel}$ as the characteristic scattering angle below which 50% of the electrons are inelastically scattered into.

The analytical model is derived based on the characteristic angles: elastic $\theta_{el}$ and inelastic $\theta_{inel}$. These angles depend on the electron energy, and they can be obtained by numerically integrating the differential cross sections (Eq. 1 and Eq. 3) azimuthally [1-4], then being normalized by the total cross section, finally summed in the altitude dimension from 0 to $\pi$ with a fine step ($\Delta\theta_{alti} = 0.001\; mrad$). As result, for both $\theta_{el}$ and $\theta_{inel}$, the angle corresponding to 50% probability of electron being scattered into, can be obtained as the characteristic angle. The characteristic angles of elastic and inelastic scattering are shown in Fig. 1a as functions of the electron energy.

The ratio of the total inelastic scattering cross-section and the total elastic scattering cross-section can be approximately expressed by Eq. 4 [2,7].

$$R_{in2el} = \frac{\sigma_{inel}}{\sigma_{el}} \approx \frac{\gamma}{Z} \tag{4}$$

Here, $\gamma$ is a parameter close to 20 and hardly dependent on the atomic number or electron energy. This relationship holds for thin samples where multiple scattering is negligible and essentially all the high angle elastic scattering is collected [2]. As an example, the ratio can be approximated to ~3 for amorphous ice ($Z \approx$ 8). One can convert the elastic and inelastic scattering cross sections to the corresponding weights in the ultimate scattering angular distribution; thus, the effective critical angle can be estimated by Eq. 5.

xiyang@bnl.gov

$$\theta_{eff}(E) = \frac{R_{in2el}}{R_{in2el}+1}\theta_{inel}(E) + \frac{1}{R_{in2el}+1}\theta_{el}(E) \tag{5}$$

Based on the critical angle of elastic (black) and inelastic (red) scattering shown in Fig. 1a, we can obtain the effective critical angle as a function of electron beam energy (Fig. 1b). For the ultimate resolution as a function of electron beam energy, one must take the following three factors into account: 1) geometrical broadening (GB) due to the semi-convergence angle and sample thickness; 2) emittance contribution (EC) to the focused beam waist size; and 3) scattering broadening (SB) due to the angular distribution induced by both single and multiple elastic and inelastic scatterings. The ultimate beam size is the quadrature sum of the contributions from GB, EC, and SB:

$$\sigma_{\text{tot}} = \sqrt{\sigma_{\text{GB}}{}^2 + \sigma_{\text{EC}}{}^2 + \sigma_{\text{SB}}{}^2}. \tag{6}$$

The transverse beam size varies according to the sample thickness. The maximum beam size could limit the ultimate resolution; thus, we treat the beam size as the resolution in the graphs shown in Fig. 2. For the beam energy of 3 MeV (blue), 10 MeV (black), and 30 MeV (red), the relations between the beam size and the sample thickness are shown in Fig. 2a and 2b with the probe beam focused on the middle and top of the sample, respectively. For the 3 MeV case, the resolution is dominated by the SB; once the sample thickness exceeds 2 μm (Fig. 2c), the resolution, which is estimated via the maximum beam size in sample thickness direction, does not depend on where the beam is focused anymore. Like Fig. 2c, a different electron beam energy of 10 MeV is shown in Fig. 2d.

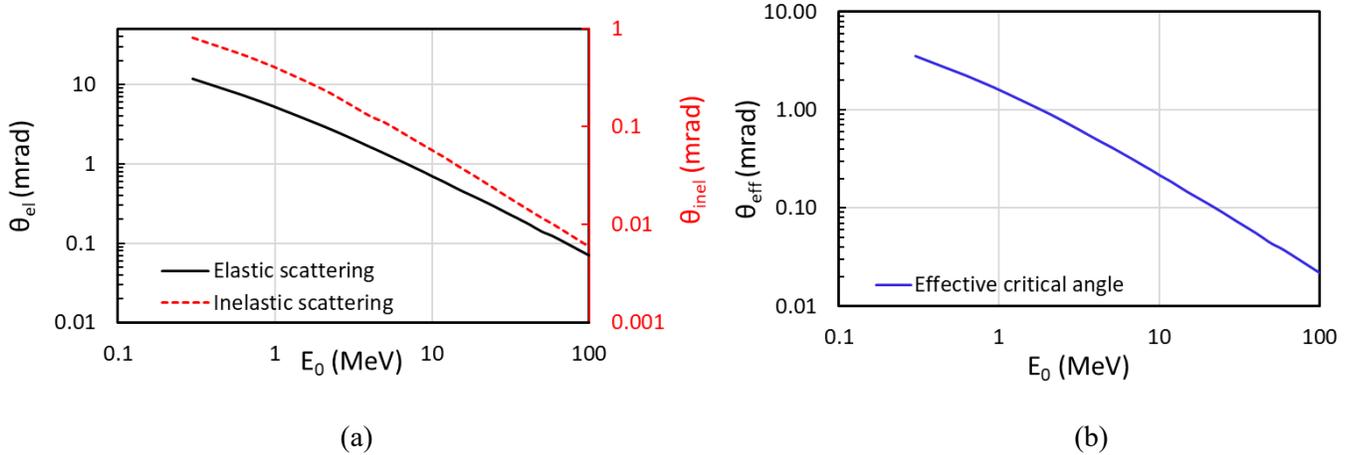

(a)                                                            (b)

Fig. 1. (a) Critical angles of elastic (black) and inelastic (red) as a function of electron beam energy are plotted. (b) The effective critical angle estimated by Eq. 5 as a function of the electron beam energy is plotted.

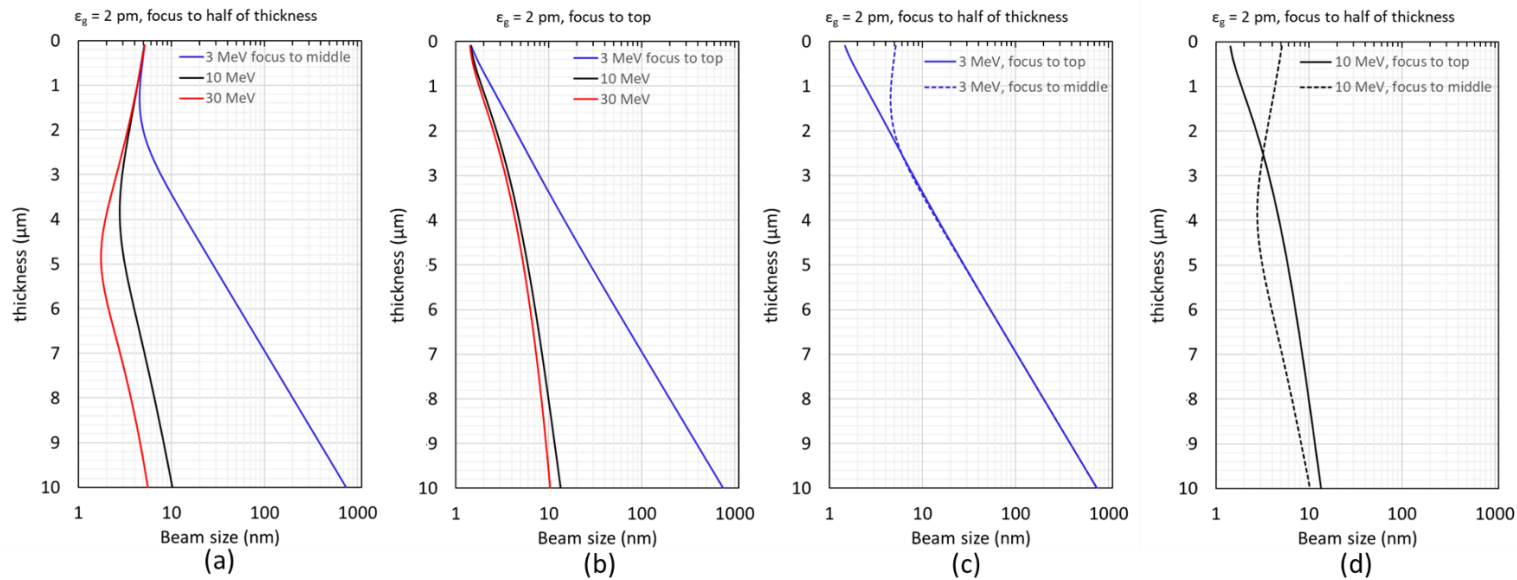

Fig. 2. Regarding 3 different beam energies, 3 MeV (blue), 10 MeV (black), and 30 MeV (red), the beam size as a function of sample thickness, when the probe beam is focused to: (a) the middle; (b) the top. (c) For the 3 MeV case only, the beam size is plotted as blue- solid and dashed curves when the beam is focused to the top and the middle, respectively. (d) Like (c) with different electron beam energy of 10 MeV.

Despite where to focus the electron beam on the sample thickness dimension (see Fig. 2a and 2b), it is required to have the electron beam energy at least 10 MeV to achieve nanoscale resolution (≤10 nm) for in-situ imaging large bio-samples with the thickness up to 10 μm. However, further increasing the electron beam energy doesn't improve the resolution anymore. It is evident in Fig. 3 that the optimal choice of electron beam energy for in-situ imaging of a large thick bio-sample is 10 MeV.

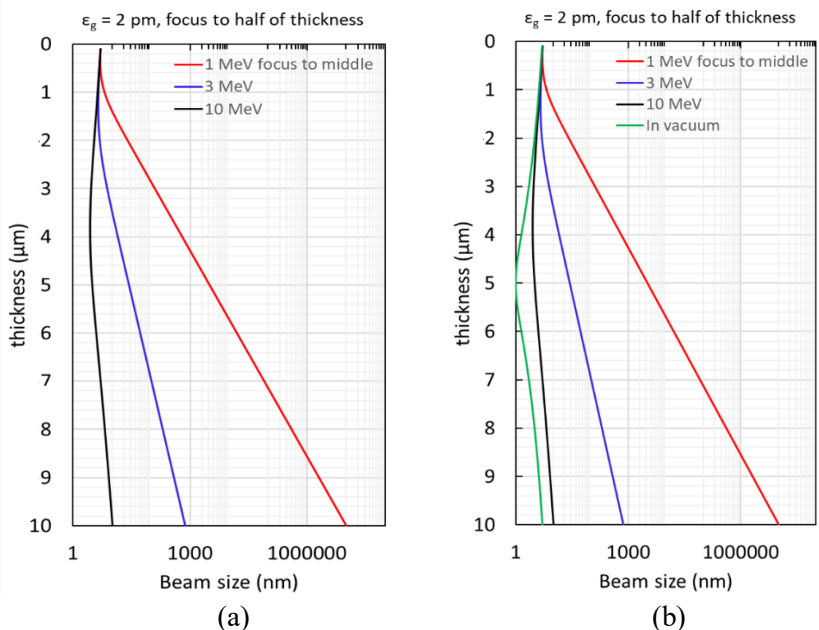

Fig. 3. Regarding 3 different beam energies, 1 MeV (red), 3 MeV (blue) and 10 MeV (black): (a) the beam size vs sample thickness; (b) their comparison to the case of in vacuum (green).

## II. Optimized Beam Energy for Different Sample Thickness

We explore the relationship between the electron beam energy and the optimal sample thickness. So far, we assumed that the tolerance of the maximum projected beam size in the sample thickness dimension is 10 nm. If the sample is thicker than 10 μm and the beam energy is higher than 10 MeV, the GB and EC could ultimately make the resolution worse than 10 nm. However, for the sample thickness up to 10 μm, the electron beam energy can be optimized, as shown in Fig. 4.

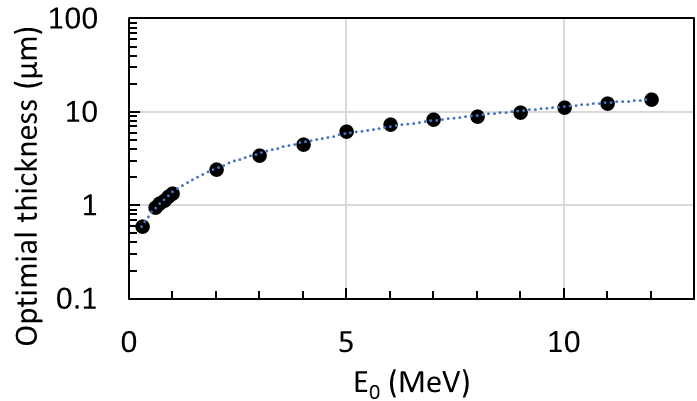

Fig. 4. Electron beam energy determines the optimal sample thickness when sample thickness is equal to or less than 10 μm.

## III. Detector Signal at Different Electron Energy

The detector signal is defined as the fraction of incident electrons traversing through the entire sample (e.g., the amorphous ice) and being collected by the detector within a certain angle range (e.g., 0 to 1 mrad), see the schematic layout in Fig. 5a. The detector signal calculated using the analytical model we implemented is shown in Fig. 5b as a function of the electron beam energy for two different detector angle ranges: 0 to 1 mrad (red) and 0 to 10 mrad (black). The normalized peak intensity on-axis at the detector is shown in Fig. 5c as a function of sample thickness for three electron beam energies: 1 MeV (red), 3 MeV (blue), and 10 MeV (green). The detector signal includes all 5 channels [1] – none interacted, single and multiple elastic and inelastic scattering electrons, within the angle of the detector ranging from 0 to 10 mrad.

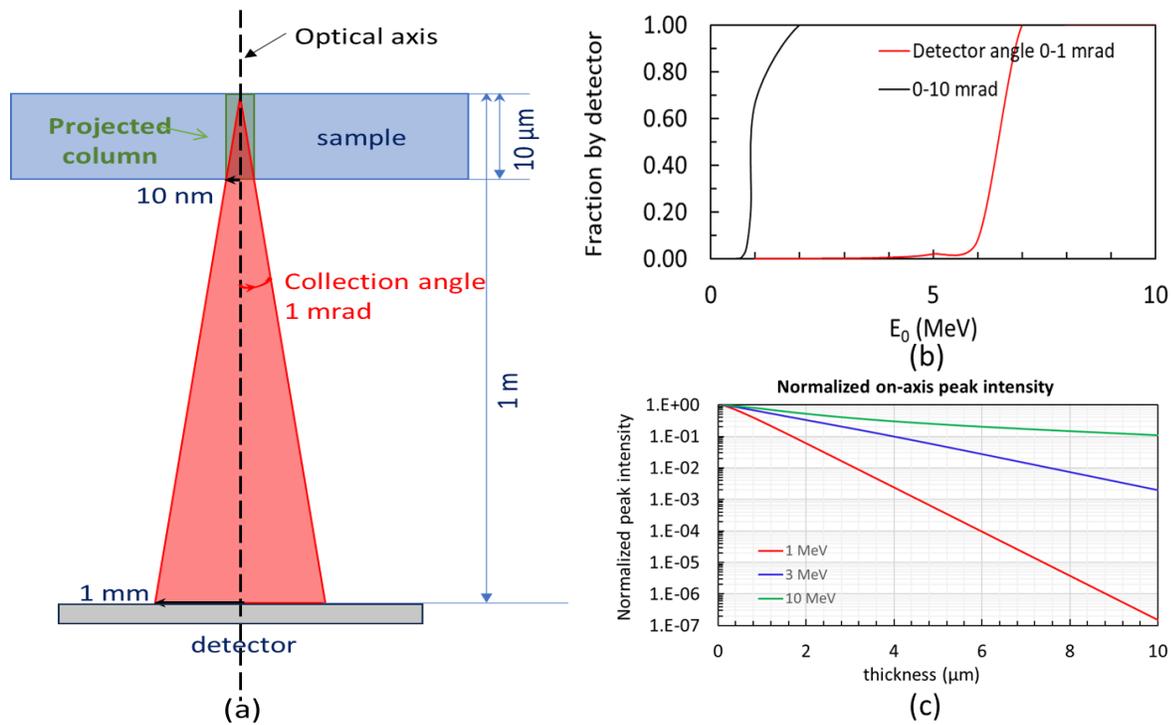

Fig. 5. (a) Schematic of the probe beam traversing from the sample to the detector (e.g., detector angle relative to the optical axis and the top surface of the sample from 0 to 1 mrad) . (b) Fraction of the collected electrons, which are normalized by the total number of incident electrons, as a function of the electron beam energy for two different detector angle ranges: 0 to 1 mrad (red) and 0 to 10 mrad (black). (c) Normalized peak intensity on-axis at the detector as a function of sample thickness is plotted at three different electron beam energies, 1 MeV (red), 3 MeV (blue), and 10 MeV (green).

**Conclusion**

We derived the analytical model based on the characteristic angles of elastic and inelastic scattering. The model is applied to explore the relationship between imaging resolution, sample thickness and beam energy. As the result, while the sample thickness ≤10 µm, there exists an optimal electron beam energy below 10 MeV regarding each specific sample thickness. However, when the sample is thicker than 10 µm, the optimal beam energy should be 10 MeV and the ultimate resolution will become worse with the increase of the sample thickness due to the geometrical broadening effect. Based on the scattering probability being proportional to the mass density, the above results can be adopted to different materials with a specific multiplier, which is the ratio of mass density between the targeting material and the amorphous ice used in the model.

## ACKNOWLEDGMENTS
We would like to thank Drs Liguo Wang, Yimei Zhu, and Lijun Wu for discussions on electron scattering in specimens.

**References:**

1. X. Yang, et al., "Towards construction of a novel nm resolution MeV-STEM for imaging of thick biological samples," submitted to Scientific Reports.
2. S. G. Wolf, *et al*., STEM Tomography in Biology, in Cellular Imaging: Electron Tomography and Related Techniques, E. Hanssen, Editor. 2018, Springer International Publishing: Cham. p. 33-60.
3. S. G. Wolf, E. Shimoni, M. Elbaum, and L. Houben, in *Cellular Imaging: Electron Tomography and Related Techniques* (ed Eric Hanssen) 33-60 (Springer International Publishing, 2018).
4. N. Tanaka, *Electron Nano-Imaging*. 1st edition, (Springer, Tokyo, 2017).
5. Du, M. & Jacobsen, C. Relative merits and limiting factors for x-ray and electron microscopy of thick, hydrated organic materials. *Ultramicroscopy* **184**, 293-309, doi:10.1016/j.ultramic.2017.10.003 (2018).
6. Reimer, L. & Kohl, H. *Transmission Electron Microscopy*. 5th edn, (Springer, New York, 2008).
7. H. Kohl and L. Reimer, Transmission Electron Microscopy. 5th ed. 2008: Springer Science+Business Media, LLC.